\documentclass[12pt]{iopart}

\usepackage{iopams}
\usepackage{amstext}
\usepackage{bbm}
\usepackage{epsfig}
\usepackage{cite}

\newcommand{\R}{\ensuremath{\mathbbm R}}
\newcommand{\N}{\ensuremath{\mathbbm N}}

\newcommand{\ket}[1]{\ensuremath{\left|#1\right\rangle}}

\newcommand{\bracket}[2]{\ensuremath{\left\langle#1\left|#2\right\rangle\right.}}

\newcommand{\avg}[1]{\ensuremath{\left\langle#1\right\rangle}}

\newcommand{\bref}[1]{(\ref{#1})}

\newcommand{\real}{\ensuremath{\mathrm{Re}}}

\newcommand{\JSP}{J.~Stat.~Phys.}
\newcommand{\PRE}{Phys.~Rev.~E}

\begin{document}

\title{Matrix product approach for the asymmetric random average process}

\author{F Zielen and A Schadschneider}

\address{Institute for Theoretical Physics,
University of Cologne, Cologne, Germany}

\eads{\mailto{fz@thp.uni-koeln.de}, \mailto{as@thp.uni-koeln.de}}

\begin{abstract}
We consider the asymmetric random average process which is a one-dimensional
stochastic lattice model with
nearest neighbour interaction but continuous and unbounded state
variables. First, the explicit functional representations,
so-called beta densities, of all local interactions leading to
steady states of product measure form are rigorously derived. This
also completes an outstanding proof given in a previous publication.
Then, we present an alternative solution
for the processes with factorized stationary states by using a
matrix product ansatz. Due to continuous state variables we obtain
a matrix algebra in form of a functional equation which can be
solved exactly.
\end{abstract}


\section{Introduction}

In recent years the study of stochastic systems has become an
attractive and important research field in modern statistical
physics. This is mainly based on the fact that in particular a lot
of interdisciplinary problems are described best by probabilistic
models. In addition, many stochastic processes represent simple
nonequilibrium systems and may serve as a kind of toy models for
the evaluation of a still outstanding nonequilibrium theory.

In this work we focus on the asymmetric random average process
(ARAP) \cite{KrugJ:asyps,RajeshR:conmm}. The model is defined on a
lattice and equipped with a probabilistic nearest neighbour
interaction. This is in common with most stochastic models,
especially with the asymmetric simple exclusion process (ASEP),
e.g., \cite{derrida:asep97} and references therein, representing
somewhat like a standard model of nonequilibrium physics. However,
the state variables of the ARAP located at the lattice sites are
continuous and unbounded while most of the models known deal with
discrete and even finite local state spaces.

Nevertheless, the ARAP is not an artificial construction. Many
physical problems are rather located in continuous than in
discrete space. E.g., it is closely related with the q-model of
granular media \cite{CoppersmithSN:forfbp}. The traffic model
of Krauss \etal \cite{krauss:krauss96} can also be mapped onto the ARAP
\cite{zielen:diss}. Furthermore, the ARAP  may show some new
phenomena undiscovered in discrete systems so far, e.g., in
\cite{zielenschad:prl} a new kind of twofoldly broken ergodicity
has been studied.

The results presented in this paper are structured as follows.

In section \ref{ch_definition} we give a definition of the ARAP in
context of the so-called quantum formalism of stochastic
processes \cite{alcaraz:stochastic94,schuetz:review}. This is useful for the application of the matrix
product ansatz (MPA) presented in section \ref{ch_mpa}.

In sections \ref{ch_pms} and \ref{ch_mpa} we focus on ARAPs with
interactions leading to factorized steady states. In
\cite{zielenschad:emfs} these interactions have been identified,
however, the calculation was not rigorously and involved a
unproven conjecture. According to this, we complete the
outstanding proof in section \ref{ch_pms}. Furthermore, we derive
an {\em explicit} functional representation of the interactions,
so-called beta densities, and discuss the influence of finite
system sizes.

Finally, in section \ref{ch_mpa} we apply the matrix product
ansatz to ARAPs with interactions identified in the previous
section. This approach has been successfully developed for quantum
spin chains and stochastic systems in the last decade (see section
\ref{ch_mpa} for references), however, it has always been applied
to systems with discrete state variables. Here we enhance its
validity to systems with continuous state spaces. The
corresponding algebras become compact functional equations with
closed solutions.

In the conclusion we discuss the MPA for ARAPs with interactions
that do not lead to product measure steady states.

\section{Definition of the process} \label{ch_definition}

\begin{figure}[ht]
  \centerline {\epsfig{file=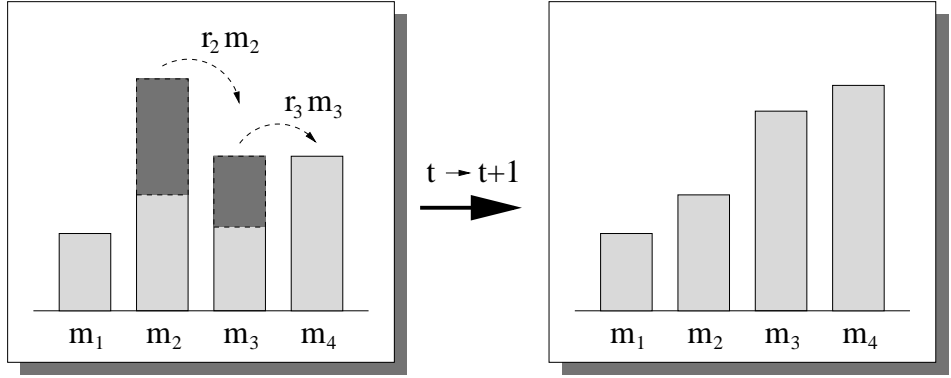, width=0.80\textwidth}}
  \caption{Mass representation of the ARAP with parallel dynamics. The height of a mass stick corresponds to $m_i$. The fragments $r_i m_i$ are shaded.}\label{modelbase_fig}
\end{figure}

The ARAP is defined on a 1D periodic lattice with $L$ sites.  Each
site $i$ carries a non-negative continuous mass variable $m_i \!\in\!
{\mathbbm R}^{+}_{0}$. The dynamics is given in discrete time and every time step $t\to t+1$ for each
site a random number $r_i \in [0,1]$ is generated from a
time-independent probability density function $\phi=\phi(r_i)$, called {\em fraction density}.
The fraction $r_i$ determines the amount of mass $\Delta_i = r_i m_i$ transported from site $i$ to
site $i+1$. The transport is completely asymmetric, i.e., no mass
moves in the opposite direction $i+1\to i$, and we obtain
\begin{equation} \label{mass_update}
  m_i \rightarrow (1-r_i) m_i + r_{i-1} m_{i-1} \;.
\end{equation}
These update rules correspond to a parallel dynamics and are illustrated in figure \ref{modelbase_fig}.

So there are only three parameters in the ARAP: first, the system size $L$
and the mass density $\rho=\frac{M}{L}$ whereby $M=\sum_i m_i$
represents the total mass. Due to mass conserving dynamics the
density is fixed and accordingly $\rho$ can be considered as a
thermodynamic variable. However, the most powerful "parameter" is
given by the fraction density $\phi$ which allows the ARAP to be
customized to a lot of interdisciplinary problems. In
\cite{zielen:diss} some stochastic models have been rewritten in
context of the ARAP by the use of suitable $\phi$-functions, e.g.,
the q-model from granular media or the Krauss-model from traffic
flow theory.

Now the ARAP will be presented in terms of the so-called quantum
  formalism for stochastic processes, e.g., used in
\cite{alcaraz:stochastic94,schuetz:review}. We consider the orthonormal state space which is spanned by the continuous
ket basis $\left\{\ket{\mathbf{m}}\right\}$ with configuration vectors $\mathbf{m}=(m_1,\ldots,m_L) $ and equipped with the inner product $\bracket{\mathbf{m}_1}{\mathbf{m}_2}
\equiv \delta\left(\mathbf{m}_1-\mathbf{m}_2\right)$ (here $\delta$ represents the
$\delta$-function). Correspondingly states of the ARAP at time $t$ are
given by

\begin{equation} \label{state}
  \ket{\mathbf{P}(t)} = \int_0^\infty \rmd^Lm P(\mathbf{m},t) \ket{\mathbf{m}}
\end{equation}
whereby $\rmd^Lm P(\mathbf{m},t)$ is a non-negative probability
measure or reworded: $P(\mathbf{m},t)$ gives the probability
density of finding the system in the configuration $\mathbf{m}$
at time $t$. Here the abbreviated form $\int_I \rmd^nx \equiv
\int_I \ldots \int_I \rmd^nx$ with $I\subset\R$ has been
introduced. Furthermore we assume $P$ to be normalized, i.e.
$\int_0^\infty \rmd^Lm P(\mathbf{m},t) = 1$.

Our main aim is to calculate the function $P$ which corresponds to the solution of the problem.
In case of the parallel dynamics given above we obtain the following $t\rightarrow
t+1$ map of the basis:
\begin{equation} \label{par_upd_basis}
  \ket{\mathbf{m}} \rightarrow \mathcal{T}\ket{\mathbf{m}} \equiv \int_0^1 \rmd^Lr \phi(\mathbf{r})
  \ket{T(\mathbf{r}) \mathbf{m}}
\end{equation}
with $\phi(\mathbf{r})\equiv\prod_i \phi(r_i)$ and
\begin{equation} \label{t_matrix}
T_{i,j}(\mathbf{r}) = (1-r_i) \delta_{i,j} + r_{i-1} \delta_{i-1,j} \;.
\end{equation}
Here $T(\mathbf{r})$ represents an $L\times L$ matrix with diagonal elements
$1-r_i$ and lower band entries $r_i$. Based on periodic boundary
conditions the top right entry is also unequal to zero, i.e.
$T_{1,L}(\mathbf{r}) = r_L$. So $\mathbf{m} \to T(\mathbf{r}) \mathbf{m}$ is nothing else
 than the compact matrix
formulation of \bref{mass_update}. Note that $T(\mathbf{r})$ operates on the space of configuration vectors $\mathbf{m}$, whereas $\mathcal{T}$ operates on the state space spanned by $\ket{\mathbf{m}}$.

Evaluating $P(\mathbf{m}',t+1) =
\bracket{\mathbf{m}'}{\mathbf{P}(t+1)} = \bracket{\mathbf{m}'}{\mathcal{T}
\; \mathbf{P}(t)}$ by using the relations
\bref{state},\bref{par_upd_basis} and \bref{t_matrix} we finally
obtain the master equation
\begin{equation}\label{mgl_par}
P(\mathbf{m}',t+1) = \int_0^\infty \rmd^Lm \int_0^1 \rmd^Lr \;
\phi(\mathbf{r}) \delta(\mathbf{m}'-T(\mathbf{r})\mathbf{m})
P(\mathbf{m},t) \;
\end{equation}
representing the fundamental dynamical equation of the ARAP.
Here the expression $\phi(\mathbf{r}) \delta(\mathbf{m}'-T(\mathbf{r})\mathbf{m})$ represents the transition
probability density from state $\ket{\mathbf{m}}$ into state $\ket{\mathbf{m}'}$.

From now on we would like to focus on steady state dynamics only,
i.e., we look for time-independent solutions $P(\mathbf{m})$ of
\bref{mgl_par}. This simplifies the master equation to the
eigenvalue problem $\ket{\mathbf{P}}=\mathcal{T} \ket{\mathbf{P}}$. For
ergodic ARAPs, defined by dynamics that allow the system to
evolve from any given initial state to any final state in a finite
time, the steady state is equal to the infinite time limit, so
$\ket{\mathbf{P}} = \lim_{t\to\infty} \ket{\mathbf{P}(t)}$.

Although we do not focus on ARAPs with discrete masses or other
kind of updates in this paper, we would like to mention that it is
straightforward to deduce the corresponding master equations from
\bref{mgl_par}. E.g., the discrete ARAP is embedded canonically by
using $\delta$-functions, whereas ARAPs with continuous time
dynamics can be treated by a suitable chosen fraction density that
interpolates between parallel and random sequential (=continuous
time) updates. Both cases are explicitly treated in
\cite{zielen:diss}.

Furthermore, we do not take into account systems with state-dependent fraction densities, e.g., given by $\phi(r,m)$ (here the
probability of mass shifts also depends on the height of the
column). Although these ARAPs may show interesting phenomena
\cite{zielenschad:prl}, an analytical treatment is more difficult
in general.

So, the ARAP with continuous state variables, discrete parallel
update and state-independent fraction density $\phi(r)$ spans an
appropriate framework for analytical research. We hope that a lot
of results can be transferred to related ARAPs in a next step.

\section{Product measure solutions} \label{ch_pms}

In this section the complete set of ARAPs with product measure solutions
\begin{equation} \label{pmsols}
P(\mathbf{m})=\prod_i P(m_i)
\end{equation}
is presented, i.e., we determine rigorously the set $\mathcal{M}$ of all $\phi$-functions leading to factorized mass distributions.
Basically, we complete the proof of \cite{zielenschad:emfs} which includes a conjecture based on high order computations so far. Furthermore, the fraction densities of $\mathcal{M}$ are given in a closed form in contrast to our previous paper \cite{zielenschad:emfs} where the $\phi$-functions are in moment representation only. Finally, we briefly discuss the influence of the system size $L$.

\subsection{Explicit form of $\mathcal{M}$}

In \cite{zielenschad:emfs} it is shown by a constructive approach
that $\mathcal{M}$ is a two parametric set of $\phi$-functions,
determined by the first and second $\phi$-moments $\mu_1$ and
$\mu_2$ which are defined by $\mu_n = \int_0^1 \rmd r \,\phi(r)  r^n$. These free parameters $\mu_1$ and $\mu_2$ have to be chosen
with respect to the general moment properties
\begin{equation} \label{parameter_space}
1>\mu_1>\mu_2\geqslant\mu_1^2
\end{equation}
only. Then the higher moments are uniquely determined by $\mu_1$ and $\mu_2$ through
\begin{equation} \label{moment_criterion}
\mu_n
= \frac{\Gamma(n+\tilde{\lambda})}{\Gamma(\tilde{\lambda})}
\frac{\Gamma(\lambda)}{\Gamma(n+\lambda)}
\end{equation}
with
\begin{equation}
\label{moment_criterion_lambda}
\tilde{\lambda} = \mu_1 \frac{\mu_1 - \mu_2}{\mu_2 - \mu_1^2} \quad \text{and} \quad
\lambda = \frac{\mu_1 - \mu_2}{\mu_2 - \mu_1^2} \;.
\end{equation}

From now on we neglect for simplicity the special case $\mu_2=\mu_1^2$ corresponding to $\phi(r)=\delta(r-\mu_1)$ and leading to product measure solutions generated by $P(m)=\delta(m-\rho)$. In particular, these singular $\delta$-densities will not be considered as elements of $\mathcal{M}$. For completeness we present the factorized mass densities associated with \bref{moment_criterion}, also deduced in \cite{zielenschad:emfs}:
\begin{equation} \label{mfsol_P}
P_\lambda(m) = \frac{\lambda^{\lambda}}{\Gamma\left( \lambda \right)}
\frac{1}{\rho} \left( \frac{m}{\rho} \right)^{\lambda-1}
\rme^{-\lambda \frac{m}{\rho}} \;.
\end{equation}

Please note that the {\em uniqueness} of the higher moments $\mu_{n\geq3}$ has already been proven in \cite{zielenschad:emfs}. Correspondingly, the densities given above span the complete set $\mathcal{M}$.
However, the explicit form \bref{moment_criterion} is a conjecture that has only been shown rigorously for $n \leq 10$. According to this, our approach in this section is canonical: We first derive explicitly the fraction densities related to \bref{moment_criterion} and after that we show that these $\phi$-functions really lead to factorized mass distributions. This will complete the outstanding proof. 

We start by rewriting \bref{moment_criterion} as a recurrence relation:
\begin{equation} \label{rec_moment_criterion}
\mu_{n+1} = \frac{n+\tilde{\lambda}}{n+\lambda} \mu_n \quad \text{with} \quad \mu_0=1 \;.
\end{equation}
Then we define the generating function
\begin{equation} \label{char_func_phi}
F(s) \equiv \sum_n \frac{1}{n!} \mu_n s^n \;.
\end{equation}
From $\mu_{n+1} \leqslant \mu_n$ we derive that $F$ is an entire function and consequently well defined. In addition, the relation
\begin{equation} \label{char_func_phi_2}
F(s) = \int_0^1 \rmd r \, \phi(r) \rme^{rs} 
\end{equation}
holds and $F(is)$ is the characteristic function of $\phi$.

From \bref{rec_moment_criterion} and \bref{char_func_phi} the differential equation
\begin{equation} \label{kummer_equ}
s F''(s) + (\lambda - s) F'(s) - \tilde{\lambda} F(s) = 0
\end{equation}
is derived after some algebra which is nothing else than Kummer's
equation \cite{abramowitz:handbook}. This ordinary differential
equation of second order is elaborated very well, e.g., it appears
in context of the hydrogen atom, and so we may rely on a huge
pool of known results \cite{abramowitz:handbook}.

For linear differential equations of second order there are always two
independent solutions. But here, only one of them, the so-called
Kummer M-function $M(\tilde{\lambda}, \lambda, s)$, is analytical
in $s=0$ as long as $\tilde{\lambda}$ is not a negative integer. This
is ensured by \bref{parameter_space} and
\bref{moment_criterion_lambda}.

Consulting \cite{abramowitz:handbook} yields for the special case
$\real \; \lambda > \real \; \tilde{\lambda}$, which is also
satisfied here, the integral representation
\begin{equation} \label{integral_rep}
M(\tilde{\lambda},\lambda,s) = \int_0^1 \rmd r \,
\frac{\Gamma(\lambda)}{\Gamma(\tilde{\lambda}) \Gamma(\lambda-\tilde{\lambda})}  r^{\tilde{\lambda}-1} (1-r)^{\lambda-\tilde{\lambda}-1}
\rme^{rs} 
\end{equation}
which finally gives an explicit representation of $\phi$ by comparison with \bref{char_func_phi_2}.

At last, this functional solution is rewritten in terms of parameters $a$
and $b$ instead of $\tilde{\lambda}$ and $\lambda$ to simplify the
representation of $\mathcal{M}$. By the transformation $a =
\tilde{\lambda}$ and $a+b = \lambda $ a symmetric form of the
fraction densities is achieved and we get
\begin{equation} \label{mcont_def}
\mathcal{M} = \left\{ \phi_{a,b}(r)=  \frac{1}{B(a,b)} r^{a-1} (1-r)^{b-1} \Bigg| a,b \in \R^+ \right\}\;.
\end{equation}
These are so-called beta densities which are very common in
probability theory \cite{feller:prob}. At this, the normalization constant is given by the beta function
\begin{equation}
B(a,b) \equiv \frac{\Gamma(a) \Gamma(b)}{\Gamma(a+b)} \;.
\end{equation}

Although the mean field models \bref{mcont_def} are parameterized by a two-dimensional manifold, the associated mass distributions \bref{mfsol_P} are connected to a one-dimensional parameter space only, i.e., several $\phi_{a,b}$ functions yield identical stationary states with $\lambda=a+b$. However, we have to keep in mind that this conclusion concerns only the steady state. The relaxation into the stationary state could differ completely.

Now we briefly discuss the product measure ARAPs \bref{mcont_def}
into more detail. 
In dependence of the fraction density the process may behave from even
critical ($a+b\to 0$) to deterministic ($a+b \to \infty$).
Here critical ARAPs are characterized by an algebraic mass decay \cite{CoppersmithSN:forfbp}. They are realized by dynamics that either shift the total mass located on a lattice site or forbid the transport.

In general, there
are three classes of $\mathcal{M}$-densities: continuous functions with
$\phi(0), \phi(1) \in \{0,1\}$, single peak functions with
$\phi(0)=\infty$ or $\phi(1)=\infty$ and double peak functions
with $\phi(0)=\phi(1)=\infty$. Only double peak functions may
result into mass densities $P(m)$ that diverge for $m\to 0$. This
reflects an almost critical behaviour because either transports of no mass ($r=0$) or the total mass ($r=1$) are preferred.
 However, all
classes may generate Gaussian-like mass densities with an
algebraic increase for small masses. Furthermore, it is
interesting that continuous and single peak densities, resp.
single and double peak densities, may lead to identical mass
distributions, whereas, this is impossible in case of continuous
and double peak ARAPs.

For $a=b=1$ we obtain the simplest version of the ARAP with uniform distribution $\phi(r)=1$. We will refer to this system as the free ARAP. In particular, there is no explicit truncation which can forbid or suppress a transport of fractions bigger than a critical value, e.g. discussed in \cite{zielenschad:prl}.

\subsection{Completion of the proof}

Now we prove that the densities \bref{mcont_def} really lead to factorized mass distributions according to \bref{mfsol_P}. In \cite{zielenschad:emfs} a simple criterion has been presented which makes it possible to determine and verify mean field solutions: Only product measure mass densities \bref{pmsols} satisfy the equation
\begin{equation} \label{cond2_Q}
F_Q(s_1,s_2) = F_Q(s_1,0) \cdot F_Q(0,s_2)
\end{equation}
with
\begin{equation} \label{mapping}
F_Q(s_1,s_2) \equiv \int_0^1 \!\! \rmd r \; \phi(r) \; Q\left(
\left(1-r\right) s_1 + r s_2 \right)
\end{equation}
and single site Laplace-transform $Q(s) = \int_0^\infty \rmd m
P(m) \rme^{-ms}$. Hence, \bref{cond2_Q} represents a sufficient
criterion for testing the validity of product measure solutions.
In the following paragraphs we will show that the mass densities \bref{mfsol_P} meet the condition \bref{cond2_Q}.

The Laplace-transform of \bref{mfsol_P} is given by
\begin{equation} \label{mfsol_Q}
Q(s) =\left(1 +\frac{\rho}{\lambda} s\right)^{-\lambda} \;.
\end{equation}
Then, using \bref{mcont_def} and \bref{mfsol_Q} with \bref{mapping} leads to
\begin{equation} \label{mapping_here}
F_Q(s_1,s_2)=  \int_0^1 \rmd r
\frac{\Gamma(a+b)}{\Gamma(a)\Gamma(b)} \frac{r^{a-1}
(1-r)^{b-1}}{\left(1+\frac{\rho}{\lambda} \left[ (1-r) s_1 + r s_2
\right] \right)^{\lambda}} \;.
\end{equation}
Evaluation of the r.h.s. of \bref{mapping_here}
can be done by use of the so-called Feynman
parameters which are a well-known tool in field theory. They are nothing else than the formula
\begin{equation} \label{feynman_parameters}
\fl\prod_{i=1}^n x_i^{-\alpha_i} = \frac{\Gamma\left(\sum_{i=1}^n
\alpha_i\right)}{\prod_{i=1}^n \Gamma(\alpha_i)} \int_0^\infty
\rmd^Lr \delta\left(\sum_{i=1}^n r_i -1\right) \prod_{i=1}^n
r_i^{\alpha_i-1} \left(\sum_{i=1}^n r_i x_i \right)^{-\sum_{i=1}^n
\alpha_i}.
\end{equation}
Here $\alpha_i$ are real and positive, whereas $x_i$ may be
complex. A derivation of \bref{feynman_parameters} is given in \cite{zinnjustin:qft}.
Now we simply apply the values
\begin{equation}
n=2\quad,\quad x_i=1+\frac{\rho}{\lambda} s_i\quad,\quad
\alpha_1=b \quad \text{and} \quad \alpha_2=a \;,
\end{equation}
integrate over $r_1$, relabel $r_2 \to r$, and finally equation
\bref{feynman_parameters} rereads as
\begin{equation} \label{feynman_arap}
\fl
\left(1+\frac{\rho}{\lambda} s_1\right)^{-b}
\left(1+\frac{\rho}{\lambda} s_2\right)^{-a} = \int_0^1 \rmd r
\frac{\Gamma(a+b)}{\Gamma(a)\Gamma(b)} \frac{r^{a-1}
(1-r)^{b-1}}{\left(1+\frac{\rho}{\lambda} \left[ (1-r) s_1 + r s_2
\right] \right)^{\lambda}} \;.
\end{equation}
Together with \bref{mapping_here} we derive
\begin{equation}
F_Q(s_1,s_2)=\left(1+\frac{\rho}{\lambda} s_1\right)^{-b}\left(1+\frac{\rho}{\lambda}s_2\right)^{-a} \;.
\end{equation}
From this follows directly the validity of condition \bref{cond2_Q} which completes the proof.

The formula \bref{feynman_parameters} has already successfully been
used in context of the q-model
\cite{CoppersmithSN:forfbp,vanleeuwen:qmodel02}. As mentioned in
the introduction this fundamental process of granular media is
strongly related with the ARAP. However, only symmetric ARAPs
defined by symmetric fraction densities $\phi(r)=\phi(1-r)$ can be
mapped onto the q-model \cite{KrugJ:asyps, zielen:diss}.
Correspondingly, the application of Feynman parameters has been
generalized to antisymmetric $\phi$-functions here.

\subsection{Finite systems}

The calculations in \cite{zielenschad:emfs} have only been focused on systems in the thermodynamic limit $L \to \infty$. But, also in case of finite systems the mean-field criterion \bref{cond2_Q} is valid and sufficient which can be shown in almost the same manner as done in \cite{zielenschad:emfs}. So ARAPs with beta densities are of product measure form irrespective of the system size $L$.

However, for $L<\infty$ the configuration space is restricted to the hyperplane
\begin{equation}
F_L(M) \equiv \left\{ \mathbf{m} \; \bigg| \;\sum_{i=1}^L m_i = M \right\}
\end{equation}
due to mass conserving dynamics. Accordingly, the mass density has to be renormalized and we obtain
\begin{equation} \label{P_finite}
P^{(L)}_\lambda(\mathbf{m}) = \left\{
\begin{array}{ll}
\frac{1}{Z} \prod_{i=1}^L P_\lambda(m_i) & \text{for} \; \mathbf{m} \in F_L(\rho L) \\
0 & \text{else}
\end{array}
\right.
\end{equation}
with
\begin{equation} \label{Z_def}
Z = Z(\lambda,\rho,L,\rho L) \equiv \int\limits_{F_L(\rho L)}
\rmd^L m \prod_{i=1}^L P_{\lambda}(m_i) = \frac{1}{\rho L}
\frac{(\lambda L)^{\lambda L}}{\Gamma(\lambda L)} \rme^{-\lambda L}
\;.
\end{equation}
Thus, projection onto the $F_L(\rho L)$ surface, i.e. fixing the total mass, corresponds to shifting our focus from grand-canonical to a canonical point of view whereby $Z$ corresponds to the canonical partition sum. A detailed calculation of $Z$ is given in \ref{Z_prove}.

One should keep in mind that the exact solutions \bref{P_finite}
are still of product measure form if restricted to $F_L(\rho L)$.
However, this coincidence with the $L=\infty$ case is only formal,
e.g., the one-site mass density is {\em not} simply given by
$Z^{-\frac{1}{L}} P(m)$. One has to take into account the
additional interaction induced by the restriction $\mathbf{m} \in
F_L(\rho L)$. So we derive by using the relation \bref{Z_relation}
again:

\begin{equation}
\fl P^{(L)}_\lambda(m_i) = \int\limits_{F_{L-1}(\rho L - m_i)}
\!\!\!\!\!\! \rmd^{L-1}\tilde{m} \; \frac{1}{Z} P_\lambda(m_i)
\prod_{j=1}^{L-1} P_\lambda(\tilde{m}_j) =
\frac{Z(\lambda,\rho,L-1,\rho L-m_i)}{Z(\lambda,\rho,L,\rho L)}
P_\lambda(m_i) \;.
\end{equation}

\section{Matrix product ansatz} \label{ch_mpa}

In this section we solve ARAPs with $\mathcal{M}$-densities by
using a matrix product ansatz (MPA). This technique has been
initially introduced for calculating exact ground states of
quantum spin chains \cite{kluemper:mpa91}. Shortly after, Derrida
and coworkers have successfully applied the MPA to a
nonequilibrium system, namely the ASEP with random sequential
dynamics \cite{derrida:mpa93}. Meanwhile the MPA has been evolved
to a standard tool for one-dimensional stochastic models, e.g.
\cite{derrida:asep97,schuetz:review,rajewsky:asep98,klauck:mpa99,hinrichsen:review00}
and references therein. However, its field of application so far is
mainly restricted to variants of the ASEP (different updates,
local defects or two species of particles).

In general the MPA is applied to systems defined on a
two-dimensional local state space, i.e., sites can be vacant or
occupied. In this case, manageable sets of algebraic
objects (corresponding to the local states) and algebraic
relations (reflecting the local dynamics) are obtained. For
example the ASEP provides one condition, $p D E = D + E$,
where the objects $E$ and $D$ correspond to holes and particles.
Nevertheless, it is rather complicated to find (matrix) representations of
that poorly defined algebra \cite{derrida:asep97}. An extension to
a model with an arbitrary, but still finite, number of local states
is given in \cite{karimipour:mpa99}.

Here we apply the MPA to a stochastic system with
\emph{continuous} state variables. As seen in the previous
section, mean field is exact for fraction densities taken from
$\mathcal{M}$, i.e., the corresponding algebras have
one-dimensional representations and are given by a functional
equation.

We derive mass solutions in agreement with the results given in
section \ref{ch_pms}. So an alternative approach for the
calculation of steady states is presented in this paper. This may
show new perspectives in the treatment of other ARAPs which are
unsolved so far. E.g., the free ARAP with continuous time update
still lacks an exact description of the steady state. Also the
state-dependent models given by mass-dependent fraction densities $\phi(r,m)$ could be treated by
the MPA.

Finally, our work extends the scope of the MPA to systems with
unbounded and continuous state spaces! Thus, we have to deal
with functions and functional equations instead of discrete
objects. For this, we may fall back on the well elaborated field
of functional theory. So, although things get in principal more
complex -- measured in degrees of freedom -- life becomes easier.

\subsection{Continuous algebra} \label{ch_algebra}

In this section we derive the matrix algebra of the ARAP in the
thermodynamic limit.

We start with the ``defect'' matrix product ansatz for
backward sequential dynamics \cite{rajewsky:asep98}. A definition
of the backward sequential update is presented in the \ref{bs}
whereas a detailed introduction to the MPA for stochastic systems
can be found in the references given in the introduction of this section. Therefore, we assume
the local interaction to obey
\begin{equation} \label{mpa_d}
t\left(A\otimes\bar{A}\right) = \bar{A}\otimes A
\end{equation}
with
\begin{equation} \label{vectors_c}
A = \int_0^\infty \rmd m D(m) \ket{m} \quad \text{and} \quad
\bar{A} = \int_0^\infty \rmd m \bar{D} (m) \ket{m} \;.
\end{equation}
Here $\ket{m}$ spans the infinite and continuous local state space of a single site and the tensor product is defined as usual, i.e.
\begin{equation} \label{dotp}
A \otimes \bar{A} = \int_0^\infty \rmd m \int_0^\infty \rmd
\tilde{m}\; D(m) \bar{D}(\tilde{m}) \ket{m,\tilde  {m}} \;,
\end{equation}
whereby we have used $\ket{m,\tilde{m}} \equiv \ket{m} \otimes \ket{\tilde{m}}$. This is in accordance with the notation introduced in section \ref{ch_definition}.
Note that the algebraic objects $D$ and $\bar{D}$ depend on a continuous parameter $m$ reflecting the mass located on a lattice site.

It is easy to see that
\begin{equation}
\ket{\mathbf{P}}_L = \tr \left( \underbrace{A
\otimes A \otimes \dots \otimes A}_{L-1 \; \text{terms}}  \otimes  \bar{A} \right)
\end{equation}
represents a steady state under
backward sequential dynamics
\begin{equation}
\mathcal{T}_\text{bs} = t_{L,1} t_{1,2} t_{2,3} \ldots t_{L-1,L} \;.
\end{equation}
The operator $t_{i,i+1}$ is defined by the local interaction $t$ acting on sites $i$ and $i+1$. For further information about update procedures please refer to \cite{rajewsky:asep98}. 

Note that trace operator
$\tr$ and time evolution operator $\mathcal{T}_\text{bs}$ commute because $\tr$ acts
on the auxiliary space of the algebraic objects whereas $\mathcal{T}_\text{bs}$ is
defined on the state space. Here the trace operator shall ensure
translational invariance of the steady state.

In the thermodynamic limit the parallel update corresponds to the
backward sequential update (see \ref{bs}), so the defect can be
neglected and we obtain
\begin{equation} \label{P_d}
\ket{\mathbf{P}} = \lim_{L \to \infty} \ket{\mathbf{P}}_L = \tr
\left( A \otimes A \otimes\dots \right) \;.
\end{equation}

Now we give the explicit definition of the local dynamics. First, we change notation (or, more formal, the basis) uniquely in the following way
\begin{equation} \label{basis_c}
\ket{m,\tilde{m}} \rightarrow \ket{m+\tilde{m},m}
\end{equation}
because the local dynamics is mass conserving. Then, the representation of the 
local interaction reads as follows:
\begin{equation} \label{dyn_par_c}
t \ket{s,m} = \int_0^m \rmd \Delta f(\Delta,m) \ket{s,m-\Delta} \;.
\end{equation}
Here $f$ represents the so-called fragment density, giving the probability (density) that a fragment of size $\Delta$ is broken of a stick with mass $m$. The densities $f$ and $\phi$ are simply related by
\begin{equation} \label{f_def}
  f(\Delta,m) = \frac{1}{m} \phi\left( \frac{\Delta}{m}\right) \;.
\end{equation}

Using \bref{mpa_d}--\bref{dotp} with \bref{basis_c}--\bref{f_def} yields after some calculation the matrix algebra in form of a functional equation:
\begin{equation} \label{algebra_c}
\int_0^{s-m} \rmd \Delta \frac{1}{s-\Delta}
\phi\left(1-\frac{m}{s-\Delta}\right) D(s-\Delta) \bar{D}(\Delta)
= \bar{D}(m) D(s-m) \;.
\end{equation}
For a detailed derivation we refer to \ref{app_alg}. Please note that we are confronted with \emph{one condition} and
\emph{two objects} only. This is remarkable due to the immense
degrees of freedom. In discrete systems every local state
corresponds to a single algebraic object, e.g., in the ASEP we
deal with an object $E$ for empty sites and $D$ for
occupied sites. Accordingly, unbounded local state spaces yield an
infinite number of algebraic objects. Additionally, the number of
conditions derived from the local dynamics and defining relations
between the algebraic objects depends quadratically on the number of
local states. So, increasing the degrees of local freedom makes it
more difficult to find representations of the algebraic objects
that fulfill all conditions. Even more fascinating is that in
case of continuous system things become easier again! The
algebraic objects condense to functions and many conditions
summarize in one functional equation.

\subsection{Constructive solution for the special case $\phi_{1,b}$}

In general $D(m)$ and $\bar{D}(m)$ are arbitrary algebraic objects
and sometimes it is possible to derive information about the
underlying system without finding a concrete representation, e.g.,
this has been done for the ASEP in \cite{derrida:mpa93} where some quantities
like the flux have been calculated by recurrently solving the
algebraic relations. Usually one tries to find a
matrix representation fulfilling the dynamical conditions. Here one
distinguishes one-dimensional and higher dimensional
representations because one-dimensional representations correspond
to product measure solutions \bref{pmsols} always, whereas higher
dimensional solutions incorporate additional correlations and show
that the process cannot be solved by a mean field ansatz.

In this subsection we will focus on ARAPs with $\phi_{1,b}$-functions, i.e. $a=1$, given by
\begin{equation}
\phi_{1,b}(r) = b (1-r)^{b-1}
\end{equation}
and part of the class $\mathcal{M}$. Without any further input from section \ref{ch_pms} we will derive the according mass densities by the matrix product technique, i.e., by solving the algebraic equation \bref{algebra_c}. Since we are looking for product measure solutions, we assume a one-dimensional representation of $D$ and $\bar{D}$. Even without knowing that such solutions exist, one would always try to solve a problem by the simplest ansatz.
 If this approach fails, we would look for higher dimensional representations.

This means that $D$ and $\bar{D}$ are nothing else than {\em functions} with one real parameter. In particular, $D$ and $\bar{D}$ commute and we assume them to be differentiable.

Using $\phi_{1,b}$ with \bref{algebra_c} we obtain
\begin{equation}
b \int_0^{s-m} \rmd \Delta  (s-\Delta)^{-b} D(s-\Delta)
\bar{D}(\Delta) = m^{1-b}\bar{D}(m) D(s-m) \;.
\end{equation}
Then differentiating
with respect to $m$ generates after rearrangement
\begin{eqnarray} \label{diff_2v_c}
(1-b) \bar{D}(m) D(s-m) + b \bar{D}(s-m) D(m) \nonumber \\
= m \bar{D} (m) D^\prime(s-m)- m \bar{D}^\prime(m)  D(s-m) \;,
\end{eqnarray}
a differential equation with two functions and two variables $s$ and $m$.
For $s=2 m $ the explicit $b$-dependence vanishes and \bref{diff_2v_c} reduces to
\begin{equation} \label{diff_1v_c}
\frac{\rmd}{\rmd m} \left[ \ln \frac{D(m)}{\bar{D}(m)} \right]  =
\frac{1}{m} \;.
\end{equation}
Directly the relation
\begin{equation} \label{rel_d_e}
D(m) = C m \bar{D}(m)
\end{equation}
is obtained. Inserting this solution in \bref{diff_2v_c} enables us to extract $D$ from the algebra and after some calculus we derive
\begin{equation} \label{rel_e}
h(m) - h(s-m) =  (b-1) \frac{s-2m}{m(s-m)}
\end{equation}
with
\begin{equation} \label{def_h}
h(z)\equiv\frac{\rmd}{\rmd z} \ln \bar{D}(z) \;.
\end{equation}
For the free ARAP, i.e. $b=1$,
the r.h.s.\ of \bref{rel_e} is equal to zero. Then $h$ has to be constant because \bref{rel_e} has to be valid for all $0\leqslant m \leqslant s < \infty$. Thus, we get $\bar{D}(m) = \tilde{C} \exp(\mu m
)$.
For arbitrary $b$ we rewrite the r.h.s.\ of \bref{rel_e} by expansion into partial fractions and achieve the difference equation
\begin{equation}
h(m) - h(s-m) = (b-1) \left( \frac{1}{m} - \frac{1}{s-m}\right) \;.
\end{equation}
Its general and unique solution is given by $h(z)=\mu+ (b-1) z^{-1}$ (at this we used the same argument as for the $b=1$ case). By the help of definition \bref{def_h} we obtain $\bar{D}(m) = \tilde{C} m^{b-1} \exp(\mu m)$ and finally $D(m) = \hat{C} m^b \exp(\mu m)$ by \bref{rel_d_e}.

In case of a one-dimensional solution the trace operator in \bref{P_d} becomes redundant and the identity $P(m)=D(m)$ is valid.
 Using the boundary conditions $\int_0^\infty \rmd m P(m) = 1$ and $\int_0^\infty \rmd m P(m) m = \rho$, we determine the constants $\hat{C}$ and $\mu$ and obtain the single-site mass density
\begin{equation} \label{mpasol}
P(m) = \frac{(1+b)^{1+b}}{\Gamma\left(1+b \right)} \frac{1}{\rho}
\left( \frac{m}{\rho} \right)^{b} \rme^{-(1+b) \frac{m}{\rho}} \;.
\end{equation}
which is in perfect accordance with \bref{mfsol_P}.

We emphasize that the matrix product ansatz offers a big advantage compared to other approaches dealing with product measure solutions: we do not have to prove the exactness of the solution \bref{mpasol}. Usually one assumes the master equation to be solved by a solution of type \bref{pmsols}, i.e., we make a so-called mean-field ansatz and look for solutions \cite{RajeshR:conmm,CoppersmithSN:forfbp,zielenschad:emfs}.
 The next step is to show that the mean-field ansatz is really correct, i.e., we have to prove that all higher correlations or joint probability densities decompose. Although in \cite{zielenschad:emfs} this extensive task has been reduced by deduction of the criterion \bref{cond2_Q}, the step of testing remains: mean-field solutions have to satisfy \bref{cond2_Q} for all $s_1$ and $s_2$.

However, in case of the matrix product technique the proof of exactness is delivered for free. The construction of the steady state ensures its exactness (see subsection \ref{ch_algebra}) and the dimensionality of the representation gives information about the correlations.

\subsection{Approach for the general case $\phi_{a,b}$}

Here the matrix algebra for beta densities with arbitrary $b$ and $a\in\N$ is considered. As in the previous section we derive the corresponding differential equations involving $D$ and $\bar{D}$, however, these become more complex and we have to treat them for each $a$ separately. We present the $a=2$ case in detail and refer to the problems in finding closed solutions for arbitrary $a$.

Starting point is \bref{algebra_c} with fraction densities \bref{mcont_def}. Introducing $H(z)\equiv z^{1-(a+b)} D(z)$ we rewrite the functional equation to
\begin{equation} \label{func2}
\fl \int_0^{s-m} \!\!\! \rmd \Delta (s-m-\Delta)^{a-1} H(s-\Delta) \bar{D}(\Delta) = B(a,b) m^{1-b} \bar{D}(m) D(s-m) \;. 
\end{equation}
Now we assume $a$ to be integer and derive the relation
\begin{equation} \label{rel_der}
\fl \frac{\partial^a}{\partial m^a} \int_0^{s-m} \!\!\! \rmd \Delta (s-m-\Delta)^{a-1} H(s-\Delta) \bar{D}(\Delta) = (-)^a (a-1)! H(m) \bar{D}(s-m) \;.
\end{equation}
For further information please consult \ref{app_abl}. So, by an $a$ times differentiation the integral expression disappears and combination of \bref{func2} and \bref{rel_der} yields the equation
\begin{equation} \label{diff_1}
\frac{\partial^a}{\partial m^a} \left[ m^{1-b} \bar{D}(m) D(s-m) \right] = (-)^a \frac{\Gamma(a+b)}{\Gamma(b)}  m^{1-(a+b)} D(m) \bar{D}(s-m) \;.
\end{equation}
For $a=1$ we obtain \bref{diff_2v_c}. Defining $G(z)\equiv z^{1-b} \bar{D}(z)$, evaluating the expression on the l.h.s. of \bref{diff_1} and setting $s=2m$ results in
\begin{equation} \label{dgl}
\sum_{k=0}^a {a \choose k} (-)^k D^{(k)} (m) G^{(a-k)} (m) = (-)^a \frac{\Gamma(a+b)}{\Gamma(b)} m^{-a} D(m) G(m) \;. 
\end{equation}
Here the upper index denotes the $k^{\text{th}}$ derivative.
Thus, equation \bref{dgl} represents a homogeneous linear differential equation of order $a$ in $D$ and $G$ respectively. In terms of $\mathbf{F}\equiv(D,G)$ it is even nonlinear.

Henceforth, one may apply the following strategy as done in the previous section: first we try to solve \bref{dgl} in $\bar{D}$, i.e., we obtain a relation $\bar{D}(m)=\mathcal{F}(D)(m)$. Then, we insert this result into \bref{diff_1}, obtain an equation with $\bar{D}$ only and try to solve it. However, this approach becomes more difficult for $a\not=1$ because the corresponding differential equations are more complex. 

As an example, we would like to discuss the case $a=2$ into detail. For this \bref{dgl} transforms into
\begin{equation} \label{dgl_2}
D(m) G''(m) - 2 D'(m) G'(m) + D''(m) G(m) = \frac{b(1+b)}{m^2} D(m) G(m) \;.
\end{equation}
To simplify \bref{dgl_2} we make use of the substitutions $d \equiv \left( \ln D \right)'$ and $g \equiv \left( \ln G \right)'$ so that a nonlinear differential equation of first order is achieved finally:
\begin{equation} \label{diff_2}
\frac{\rmd}{\rmd m} \left[ d(m) + g(m) \right] + \left[ d(m) - g(m) \right]^2 = \frac{b(1+b)}{m^2} \;.  
\end{equation}
Unfortunately, we are not able to give a general solution of \bref{diff_2} representing a Riccati equation in $a \equiv d-g$ \cite{diffbook}. This is based on the fact that $g$ is also an unknown function. Therefore, we cannot apply elaborated solution schemes for this type of equation here \cite{diffbook}.

Nevertheless, we are able to present a special solution of \bref{diff_2} given by
\begin{equation} \label{ans_sol}
d(m)=\frac{\kappa}{m}+\mu \quad \text{and} \quad g(m)=\frac{\bar{\kappa}}{m}+\mu 
\end{equation}
with $(\kappa-\bar{\kappa})^2 -(\kappa+\bar{\kappa}) = b(1+b)$. Going back, exact expressions for $D$ and $\bar{D}$ are derived, especially $D(m) = C m^{\kappa} \exp(\mu m)$. Although these functions do not fulfill the overall condition \bref{diff_1} for $a=2$, this heuristic approach yields the correct form of $D$ (see below), but an erroneous defect term $\bar{D}$.

However, we have not exploited the relations between $D$ and $\bar{D}$ to the full. According to this, we are confronted with three parameters $C$,$\kappa$ and $\mu$ now, whereas in the previous section only two unknown constants occurred. To avoid this situation, the second mass moment
\begin{equation}
\avg{m^2}= \int_0^\infty \rmd m P(m) m^2 = \frac{\mu_1(1-\mu_1)}{\mu_1-\mu_2} \rho^2
\end{equation}
is used as a third condition. It has been calculated exactly in \cite{KrugJ:asyps}. With a final reassignment $D \to P$ the solution
\begin{equation}
P(m) = \frac{(2+b)^{(2+b)}}{\Gamma(2+b)} \frac{1}{\rho} \left( \frac{m}{\rho} \right)^{1+b} \rme^{-(2+b)\frac{m}{\rho}} 
\end{equation}
is obtained which is in perfect accordance with \bref{mfsol_P}.

So the case $a=2$ differs completely from the case $a=1$ presented in the previous section: it seems to be impossible to derive both, $D$ and $\bar{D}$ from the algebraic equation \bref{algebra_c}. We have to keep in mind that the defect term $\bar{D}$ is merely an auxiliary construct and not part of the final solution \bref{P_d}. Accordingly, there could be some freedom of choice for the defect. E.g., a lot of matrix product solutions for stochastic systems are characterized by an a priori choice of the defect terms \cite{derrida:asep97,rajewsky:asep98,hinrichsen:review00}. This simplifies the algebra and leads to the proper solutions. In this way, \bref{ans_sol} represents a suitable choice of the defect function $g$.

In principle, one could try to solve the algebra for $a=3,4,\ldots$ explicitly and propose the general form \bref{mfsol_P} for arbitrary $a$. But this approach is discontenting. On the one hand, we do not know about the complexity and resolvability for increasing integer $a$. On the other hand, we do not provide a method of resolution for not-integer $a$. Finally, we are still lacking of an enclosing MPA calculus that is valid for all $\mathcal{M}$ densities. However, the MPA might nevertheless be useful for special ARAPs with non factorizing mass densities.

\section{Summary and Outlook}

In this paper we have focused on the asymmetric random average process under parallel dynamics. In the first part (section \ref{ch_pms}) it has been rigorously proven that only ARAPs equipped with beta densities and $\delta$-densities lead to product measure solutions. Here, our calculations have completed a derivation given previously in \cite{zielenschad:emfs}. In a second part (section \ref{ch_mpa}) we have presented an alternative approach, a matrix product ansatz, for deriving the mass densities under use of beta densities. Restricted to fraction densities $\phi_{1,b}$ we have given a complete calculation of the associated mass densities. For the remaining densities $\phi_{a,b}$ with $a\not=1$ we have outlined an approach based on differential equations. Altogether, we have extended the scope of the MPA to systems with continuous and unbounded state variables.

However, due to the restriction to ARAPs in class $\mathcal{M}$  we deal with a one-dimensional representation of the algebraic objects $D$ and $\bar{D}$ only. In a next step, we should concentrate on ARAPs with non-vanishing correlations which is associated with higher dimensional representations of the algebraic objects. Especially, we believe that the MPA could be the appropriate tool to solve the free ARAP with $\phi(r)=1$ under continuous time dynamics. Although this process is given by a rather simple fraction density it does not belong to the class of mean-field models \cite{KrugJ:asyps,RajeshR:conmm}. Furthermore, ARAPs with state-dependent fraction densities $\phi(r,m)$ could be treated with the MPA. This could bring forward the analysis of truncated processes. On the other hand, we could ask for $\phi$-functions that lead to solvable algebras, i.e., solvable functional equations.

In addition, the question arises if the algebraic objects of correlated ARAPs are always representable by matrices, i.e. $D(m) =\left( D_{ij}(m) \right)$. Or is it even possible to construct ARAPs that are solved by algebraic objects with continuous representations? Finally, one could identify the class of ARAPs whose stationary states may be written as matrix products. This would be very similar to \cite{klauck:mpa99} where it has been shown that stochastic processes with a finite number of local states, a finite range of interaction and continuous time dynamics are (formally) always solved by MPA.

\ack

We like to thank J H Snoeijer and J M J van Leeuwen for
very interesting discussions, especially the reference to the
asymmetric representation of the Feynman parameters
has been very useful. We also like to thank B\'alint T\'oth, Jesus Garc\'{\i}a, Pablo Ferrari and Gunter Sch\"utz for
helpful discussions.
A Schadschneider thanks the DAAD for support and the Institute for Mathematics
and Statistics at the University of Sao Paulo for its hospitality 
during the final stages of this work.

\appendix
 
\section{Partition sum $\boldsymbol{Z}$} \label{Z_prove}

We present a proof of the relation
\begin{equation} \label{Z_relation}
Z(\lambda, L, \rho, M) \equiv \int\limits_{F_L(M)} d^Lm \prod_{i=1}^L P^\rho_\lambda(m_i) =
\frac{1}{M\; \Gamma(\lambda L)} \left(\lambda \frac{M}{\rho} \right)^{\lambda L} \rme^{-\lambda \frac{M}{\rho} }\;.
\end{equation}
The superscript in $P^\rho_\lambda$ reminds of the $\rho$
dependence. The following calculation uses the Fourier representation of
the $\delta$-function and the relation between \bref{mfsol_P} and
\bref{mfsol_Q}, i.e., the principles of Laplace transformation:
\begin{eqnarray}
Z(\lambda, L, \rho, M) & = & \int_0^\infty \rmd^L m \prod_j P^\rho_\lambda(m_j) \delta(M-\sum_j m_j) \nonumber \\
& = & \frac{1}{2\pi} \int_{-\infty}^{\infty} \rmd p \; \rme^{\rmi p M} \prod_j Q^\rho_\lambda(\rmi p) \nonumber  \\
& = & \frac{1}{2\pi} \int_{-\infty}^{\infty} \rmd p \; \rme^{\rmi p M} Q^{\rho L}_{\lambda L}(\rmi p) \nonumber \\
& \stackrel{s=\rmi p}{=} & P^{\rho L}_{\lambda L} (M) \nonumber
\end{eqnarray}
So the r.h.s.\@ of \bref{Z_relation} is nothing else than a single site mass density \bref{mfsol_P} for the total mass $M$ that is rescaled  according to $\rho \to \rho L$ and $\lambda \to \lambda L$.

\section{Backward sequential update} \label{bs}

In the backward sequential update one starts updating an arbitrary
pair of sites ($i,i+1$) and applies the local update rules,
\begin{equation} \label{local_update}
  m_i \rightarrow (1-r_i) m_i \quad \text{and} \quad m_{i+1} \rightarrow
  m_{i+1} + r_i m_i \;,
\end{equation}
from right to left, i.e., opposite to the direction of the mass transport,
under consideration of periodicity until
reaching the initial pair. Thus, for finite $L$ the backward
sequential update differs slightly from parallel dynamics because
the last pair of sites $(i+1,i+2)$ to be updated holds the site
$i+1$ that has been updated already. However, in the infinite
system $L\to\infty$ this "error" vanishes because the difference
between both update procedures is local and not extensive with
$L$. Correspondingly the parallel update is equivalent to the
backward sequential dynamics in the thermodynamic limit.

\section{Derivation of the algebra} \label{app_alg}

Putting \bref{mpa_d}--\bref{dotp} and \bref{basis_c}--\bref{f_def} together we obtain the expression
\begin{equation} \label{alg_1}
\fl t(A\otimes\bar{A}) = \int_0^\infty \rmd s \int_0^s \rmd m \int_0^m \rmd \Delta \; \frac{1}{m} \phi\left(\frac{\Delta}{m}\right) D(m) \bar{D}(s-m) \ket{s,m-\Delta} \;.
\end{equation}
We introduce the linear substitution $x=s-m$ and $y=m-\Delta$ that comes along with the trivial Jacobian determinant $-1$. Therefore, the transformation of \bref{alg_1} becomes rather simple:
\begin{equation} \label{alg_2}
\fl t(A\otimes\bar{A}) = \int_0^\infty \rmd s \int_0^s \rmd y \int_0^{s-y} \rmd x \; \frac{1}{s-x} \phi\left(\frac{s-x-y}{s-x}\right) D(s-x) \bar{D}(x) \ket{s,y} \;.
\end{equation}
Here the change of variables generates a structure according to $\int_0^\infty \rmd s \int_0^s \rmd y \ldots \ket{s,y}$. Since
 $\ket{s,y}$ are linearly independent we are able to reduce the condition \bref{mpa_d} to the functional equation \bref{algebra_c}.

\section{Derivation of relation \bref{rel_der}} \label{app_abl}

The essential component of the calculation is the formula
\begin{equation} \label{formel}
\frac{\rmd}{\rmd x} \int_0^x \rmd y f(x,y) = f(x,x) + \int_0^x \rmd y \frac{\partial}{\partial x} f (x,y) 
\end{equation}
which can be verified easily.

By the help of \bref{formel} it is straightforward to prove \bref{rel_der} by induction. In particular, the correspondent term of $f(x,x)$ reduces to zero for $a>1$. However, for $a \not\in \N$ induction finally leads to the case $0<a<1$ which cannot be treated by \bref{formel} because the counterpart of $f(x,x)$, i.e., basically $(s-m-\Delta)^{a-1}$ at $\Delta=s-m$, is divergent here. Therefore, we have restricted to $a$ of integer form.


\Bibliography{99}

\bibitem{KrugJ:asyps}
Krug J and Garc\'{\i}a J 2000 {\it \JSP} {\bf 99} 31

\bibitem{RajeshR:conmm}
Rajesh R and Majumdar S N 2000 {\it \JSP} {\bf 99} 943

\bibitem{derrida:asep97}
Derrida B and Evans M R 1997 {\it Nonequilibrium Statistical
Mechanics in One Dimension} ed V Privman (Cambridge University
Press) p 277

\bibitem{CoppersmithSN:forfbp}
Coppersmith S N, Liu C-H, Majumdar S, Narayan O and Witten T A 1996 {\it \PRE} {\bf 53} 4673

\bibitem{krauss:krauss96}
Krauss S, Wagner P and Gawron C 1996 {\it \PRE} {\bf 54} 3707

\bibitem{zielen:diss}
Zielen F 2002 {\it Asymmetric Random Average Processes} phd thesis
(Institute for Theoretical Physics, University of Cologne)

\bibitem{zielenschad:prl}
Zielen F and Schadschneider A 2002 {\it \PRL} {\bf 89} 090601

\bibitem{alcaraz:stochastic94}
Alcaraz F C, Droz M, Henkel M and Rittenberg V  1994 {\it Ann.~Phys.} {\bf 230} 250

\bibitem{schuetz:review}
Sch{\"u}tz G M 2000 
{\it Phase Transitions and Critical Phenomena} vol~19
ed C Domb and J Lebowitz (Academic Press, Inc.)


\bibitem{zielenschad:emfs}
Zielen F and Schadschneider A 2002 {\it \JSP} {\bf 106} 173

\bibitem{abramowitz:handbook}
Abramowitz M and Stegun I A 1964 {\it Handbook of Mathematical
Functions} (New York: Dover Publications) ch 13

\bibitem{feller:prob}
Feller W 1968 {\it An Introduction to Probability Theory and its
Applications} vol 1 (John Wiley \& Sons)

\bibitem{zinnjustin:qft}
Zinn-Justin J 1996 {\it Quantum Field Theory and Critical
Phenomena} (Oxford: Clarendon Press) p 235

\bibitem{vanleeuwen:qmodel02}
Snoeijer J H and van Leeuwen J M J 2002
{\it \PRE} {\bf 65} 051306
 
\bibitem{kluemper:mpa91}
Kl\"umper A,  Schadschneider A and Zittartz J 1991 {\it J.\ Phys.\
A} {\bf 24} L955
\nonum
Kl\"umper A,  Schadschneider A and Zittartz J 1993
{\it Europhys. Lett.} {\bf 24} 293

\bibitem{derrida:mpa93}
Derrida B, Evans M R, Hakim V and Pasquier V 1993 {\it J.\ Phys.\ A} {\bf 26} 1493

\bibitem{rajewsky:asep98}
Rajewsky N, Santen L, Schadschneider A and Schreckenberg M 1998 {\it \JSP} {\bf 92} 151

\bibitem{klauck:mpa99}
Klauck K and Schadschneider A 1999 {\it Physica\ A} {\bf 271} 102

\bibitem{hinrichsen:review00}
Hinrichsen H 2000 {\it Adv.\ Phys.} {\bf 49} 815

\bibitem{karimipour:mpa99}
Karimipour V 1999 {\it Europhys.\ Lett.} {\bf 47} 304

\bibitem{diffbook}
Kamke E 1983 {\it Differentialgleichungen} vol~1 (Stuttgart: B.G. Teubner)

\endbib
\end{document}